\documentclass{evn2004}                     
\usepackage{txfonts}
\usepackage{graphicx} 
\begin{document}
   \title{Millimetre\,-\,VLBI Monitoring of AGN with Sub\,-\,milliarcsecond Resolution}

   \author{A. Pagels\inst{1}, T.P. Krichbaum\inst{1}, D.A. Graham\inst{1}, W. Alef\inst{1}, M. Kadler\inst{1}, A. Kraus\inst{1}, J.
   Klare\inst{1}, A. Witzel\inst{1},      
   J.A. Zensus\inst{1}, A. Greve\inst{2}, M. Grewing\inst{2}, R. Booth\inst{3}, and J. Conway\inst{3}
          }
\authorrunning{A. Pagels \rm et al.}

   \institute{Max-Planck-Institut f\"ur Radioastronomie, Auf dem H\"ugel 69, 53121
   Bonn, Germany
   \and
   Institute de Radioastronomie Millim\'etrique, 300 Rue de la Piscine, F-38406 St Martin d'H\`eres, France
   \and
   Onsala Space Observatory, 439 92 Onsala, Sweden
   }

   \abstract{
   Global millimetre VLBI allows detailed studies of the most central jet regions
of AGN with unprecedent spatial resolution of a few 100--1000 Schwartzschild radii to be made. 
Study of these regions will help to answer the question how the highly relativistic AGN jets
are launched and collimated. Since the early 1990s, bright mm-sources have been observed with global 3\,mm VLBI.
Here we present new images from an ongoing systematic analysis of the available observations. In
particular, we focus on the structure and structural evolution of the best observed AGN jets, taking 3C\,454.3 as a characteristic example. This
core-dominated and highly variable quasar
 shows a complex morphology with individual jet components accelerating superluminally
towards the outer structure. We briefly discuss the X-ray properties of 3C\,454.3 and present its radio- to X-ray large-scale brightness distribution.

   }

   \maketitle
%

\section{Introduction}
After demonstration of the technical feasibility of Very Long Baseline Interferometry (VLBI)
at the short wavelength of 3\,mm (\cite{readhead})
in the early 1980s, it took about 10 years, before regular and
systematic VLBI observations at this wavelength were performed.
During 1993--1996 global 3\,mm VLBI observations were organised and performed on an
ad hoc basis, with a limited number of 3-7 participating
antennas. In the second half of the 1990s, the Coordinated
Millimetre VLBI Array (CMVA) was established, which facilitated
larger global 3\,mm\,VLBI experiments and gave easier user access.
In early 1997, the first Very Long Baseline Array (VLBA) stations joined the global 3\,mm VLBI
observations. With the subsequent addition of more VLBA antennas,
the CMVA improved its performance and the resulting images got
continuously better. In 2002 the funding of the CMVA stopped,
but at the same time the IRAM interferometer on Plateau de Bure became
available as a phased array for mm-VLBI. To take advantage of
the superior sensitivity of the IRAM instruments and continuing major investments at this and other observatories, the
Global 3\,mm VLBI Array (GMVA) was founded as a successor of the CMVA ({\tt http://web.haystack.mit.edu/cmva/}).

The GMVA started operation in 2003, and now combines all
available antennas of the VLBA (eight in 2004) with the following stations in
Europe: Effelsberg, Pico Veleta, Plateau de Bure, Onsala and
Mets\"ahovi. It is hoped that in the near future more antennas
will join the GMVA. The present GMVA is 3-4 times more sensitive
than the VLBA alone, now allowing detailed studies of a much larger
number of compact radio sources than before to be made (\cite{agudo}).

\begin{table*}[htb]
\begin{center}
\caption{\label{tab:antennas} $\lambda$\,3\,mm VLBI array used at the observations of 3C\,454.3 -- Antenna Characteristics}
\begin{tabular}{llllcccccc}
\hline
\hline
Name         & $D^{\rm a}$ 		& $G^{\rm b}$ 		 & $T_{\rm sys}$$^{\rm c}$   &         &	    & Sessions$^{\rm d}$ &	 &	   &  	    \\
             & [m] 		&[K/Jy]  	 & [K]          & 1993.26 & 1994.01 &  1996.78 & 1997.28 & 1999.30 & 1999.81 \\ 
\hline
Effelsberg   & 100 		&  0.14         &   130         & $\surd$ & $\surd$ & $\surd$  & $\surd$ & $\surd$ & $\surd$ \\
Onsala       & 20  		&  0.05         &   300         & $\surd$ & $\surd$ & $\surd$  & $\surd$ & $\surd$ & $\surd$ \\
Pico\,Veleta & 30  		&  0.14         &   180         & $\surd$ & $\surd$ & $\surd$  & $\surd$ & $\surd$ & $\surd$ \\
Kitt\,Peak   & 12  		&  0.02         &   220         &   	  &         & $\surd$  & $\surd$ & $\surd$ & $\surd$ \\
Quabbin      & 14  		&  0.02         &   220         & $\surd$ &  	    & $\surd$  &   	 &   	   & $\surd$ \\
Ovro         & 6 $\times$ 10.4  &  0.02         &   500         &  	  &  	    & $\surd$  & $\surd$ & $\surd$ & $\surd$ \\
Plateau de Bure&15      	&  0.18		&   180	        &  	  &  	    &	       & $\surd$ &   	   &         \\
Hat\,Creek   & 9 $\times$ 6.1   &  0.01         &   300  	&  	  &  	    & $\surd$  & $\surd$ & $\surd$ & $\surd$ \\
Haystack     &37   		&  0.05         &   250         & $\surd$ & $\surd$ & $\surd$  & $\surd$ & $\surd$ &         \\
Mets\"ahovi  &14   		&  0.02         &   350         &  	  &  	    &	       & $\surd$ & $\surd$ &         \\
Sest         &15   		&  0.05         &   250         &  	  &  	    &	       & $\surd$ & $\surd$ &         \\
VLBA (several)&25   		&  0.28         &   150         &  	  &  	    &	       & $\surd$ & $\surd$ & $\surd$ \\
\hline 
\hline
\end{tabular}
\end{center}
\scriptsize {$^{\rm a}$ is the diameter in metres, $^{\rm b}$ is the Sensitivity in [K/Jy], $^{\rm c}$ the typical single-sideband system temperature in [K], $^{\rm d}$ sessions in with the telescop participates}
\end{table*}

With an angular resolution of up to 50\,$\mu$as, global 3\,mm VLBI
observations allow imaging the innermost regions of AGN and their
emanating jets.  With such high angular (and spatial) resolution and via a frequent monitoring of the variability, one can hope to learn more about the origin
of jets and their `driving engine'. 

\section{The Observations}
The $\lambda$\,3\,mm observations presented here were performed at 6 epochs different numbers of
antennas (see \ref{tab:antennas}).
All the data were recorded with the `MKIII' VLBI recording system and were correlated in Bonn or Haystack. The
output from the correlator was fringe-fitted with the HOPS-package ({\tt http://web.haystack.mit.edu/vlbi/hops.html}) and with the standard correlator software 
at the MPIfR. Hybrid maps could be produced using the intercontinental baselines to achieve a resolution of
50\,$\mu$as. The amplitude calibration was done in the standard manner using frequent system temperature, gain, and opacity measurements and by 
applying opacity corrected antenna
temperature measurements of the source flux density obtained at Pico Veleta and Effelsberg. 
The hybrid mapping and model fitting was done in the DIFMAP package.

Our sample of mm-bright AGN presently consist of 24 sources\footnote{B0234+285 (4C\,28.07), B0316+413 (3C\,84), B0355+508 (NRAO\,150) B0420--014, B0528+134 (OG\,147), B0607--157, B0827+243, B0851+202 (OJ\,287), B0923+292 (4C\,39.25), B1156+295 (4C\,29.45), B1226+023 (3C\,273B), B1228+12 (3C\,274), B1546+027 (OR\,178), B1611+343 (OS\,319), B1633+382 (4C\,38.41), B1638+398 (NRAO\,512), B1641+399 (3C\,345), B1957+405 (Cyg\,A), B2005+403, B2145+067 (4C\,06.69), B2200+420 (BL\,Lac), B2201+315 (4C\,31.63), B2230+114 (CTA\,102), B2251+158 (3C\,454.3)}. 
Some of these sources were observed only in snap-shot mode, with only a few VLBI scans. Other sources were observed 
with full $(u,v)$-coverage. Many of the sources were observed repeatedly during 1993--1999. Results of the observations of 1993 are published by Lobanov et al. (2002). Preliminary results for some other sources were also published in the `Millimeter-VLBI Science Workshop' (\cite{mm}) and the '2$^{\rm nd}$ Millimeter-VLBI Science Workshop' (\cite{2mm}). A more detailed
discussion of the individual sources will follow in subsequent papers.

  \begin{figure}[h!]
   \centering
   \includegraphics[width=5.5cm]{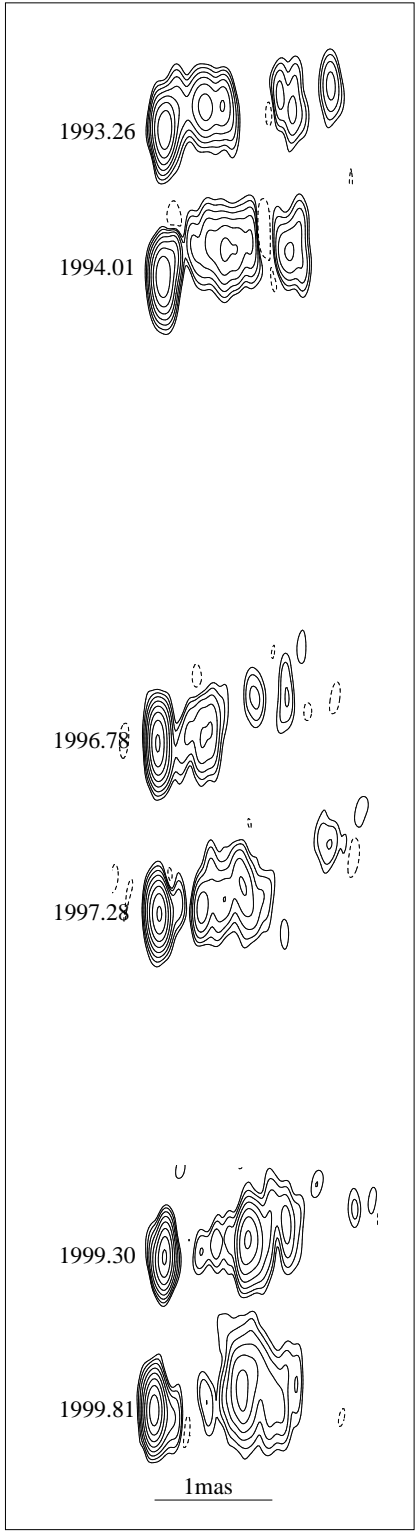}
   \caption{Global VLBI images at 86\,GHz of 3C\,454.3. The maps are restored with a common beam size of 0.28$\times$0.07\,mas at a P.A. of 0$^{\circ}$. 1\,mas corresponds to 7.7\,pc. Contour levels are: $-0.5$, 0.5, 1, 2, 4, 8, 16, 32, 64 \% of the peak flux density
           }
	   \label{fig:epochs}
    \end{figure}
 
\section{3C\,454.3}
Here we focus our analysis on the source 3C\,454.3. In the first subsection we present the results of mm-wavelengths observations, in particular at 43\,GHz and 86\,GHz. In the second subsection, we link our results in the $\mu$as-regime to the X-ray properties of 3C\,454.3 and present its radio- to X-ray large-scale brightness distribution. 

\subsection{mm-Wavelengths Observations} 
The Optically Violent Variable  3C\,454.3 is a core-dominated, highly active,
superluminal radio source at a redshift of $z$\,=\,0.859. At this redshift 1 milliarcsecond correspond
to 7.7 parsec\footnote{$H_0$\,=\,71km\,s$^{-1}$\,Mpc$^{-1}$, $\Omega_M$\,=\,0.27, $\Omega_{\Lambda}$\,=\,0.73}. Owing to its
activity and structural complexity it is one of the prime candidates for high resolution imaging (e.g.,
\cite{krichbaum}, \cite{jorstad}). VLBI images at longer cm-wavelengths show a pronounced and complex core-jet 
structure extending to the west and bending to the north (\cite{toth}).

Multi-epoch VLBA monitoring at 43\,GHz and 22\,GHz (\cite{jorstad}) report a stationary component at a core distance of
$\sim$0.6\,mas. Emission in this region was seen already earlier by Pauliny-Toth (\cite{pauliny}). The flux density of the stationary component 
varies between 3.2\,Jy (1995.31) and 1.2\,Jy (1997.19).
Observations between 1995.01 and 1997.58 indicate that three jet components were ejected. The ejection direction seems to vary between a position angle range of
$-74^{\circ}$ to $-88^{\circ}$ relative to the VLBI core.

The inner jet (r $\simeq$ 2\,mas) is oriented mainly to the west. Further out, the jet bends to the north and extends up to 10\,mas
(P.A.$= -95^{\circ}$). From early 3mm-VLBI images obtained in 1993 and 1994, 
Krichbaum et al. (1996) reported detection of apparent
superluminal motion of two inner jet components with apparent speeds of $\beta_{\rm app}\simeq$7--8\,c
at r $\le$ 1\,mas. This correspond to the proper motion measured by Pauliny-Toth et al. (1998) who found apparent velocities of $\beta_{\rm app}\simeq$ 8c at r $\ge$ 5\,mas with VLBI at cm-wavelengths\footnote{all values were recalculated for $H_0$\,=\,71km\,s$^{-1}$\,Mpc$^{-1}$, $\Omega_M$\,=\,0.27, $\Omega_{\Lambda}$\,=\,0.73}.
In a geometric interpretation this behaviour can be explained as an ultra-relativistic flow moving along a spatially bent path.

One of the best observed sources in the sample reported above is the quasar  3C\,454.3, which is one of the brightest Active Galactic Nuclei 
at 3\,mm wavelengths. It was observed with global 3\,mm-VLBI at the following epochs:
1993.26, 1994.01, 1996.78, 1997.28, 1999.30, 1999.81. 
In the following we present our results of this quasar and combine it with observations from other wavelengths.

We present in Fig.\ref{fig:epochs} six 3\,mm VLBI maps of 3C\,454.3. The source shows a one--sided core jet structure with an ultra-compact core ($\le$0.07mas) and several distinct emission components located to the west. On the mas-scale, we derive apparent
velocities of $\beta_{\rm app}= 3.7-7$\,c. This is in good agreement with the expected jet velocity and the picture
of an outward accelerating jet as suggested by  Pauliny-Toth et al. (1998). 
During our observing interval and on sub-mas scales, we observe 
even a larger change of the jet ejection angle near the core from --61$^{\circ}$ in 1993.26 to --123$^{\circ}$ in 1999.81.
In contrast to the identification scheme proposed by Jorstad et al. (2001), we do not see evidence
for a stationary component. This, however, depends strongly of the cross-identification of jet components
between the epochs, and more data need to be added to obtain a clearer picture.
Due to the limited time sampling of the 3\,mm-images, it is not yet possible 
to trace all the jet components unambiguously over time ranges larger than about 2-3 years. The addition of images
obtained at other frequencies (15 -- 43\,GHz) will help to close these gaps and we should be able to
obtain an unambiguous component identification, as our data analysis proceeds. 

   \begin{figure}[htb]
   \centering
   \includegraphics[clip,width=\columnwidth]{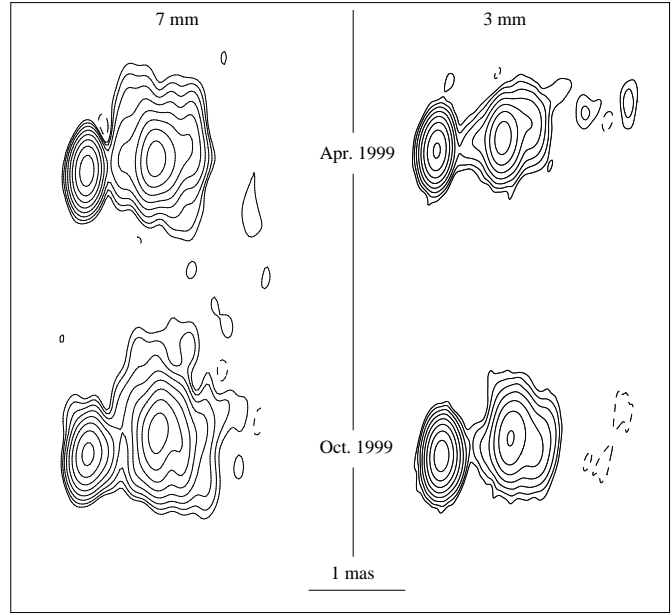}
   \caption{Comparison between $\lambda$7\,mm (left side)[A. Marscher, priv.comm] and $\lambda$3\,mm imaging (right side). All maps are convolved with a beam of
   the size 0.36$\times$0.16\,mas at a P.A. of 3.5$^{\circ}$. Contour levels are: $-0.5$, 0.5, 1, 2, 4, 8, 16, 32, 64 \% of the peak flux density
           }
	   \label{fig:7mm}
    \end{figure}
 
\begin{figure}[htb]
   \centering
   \includegraphics[clip,width=\columnwidth]{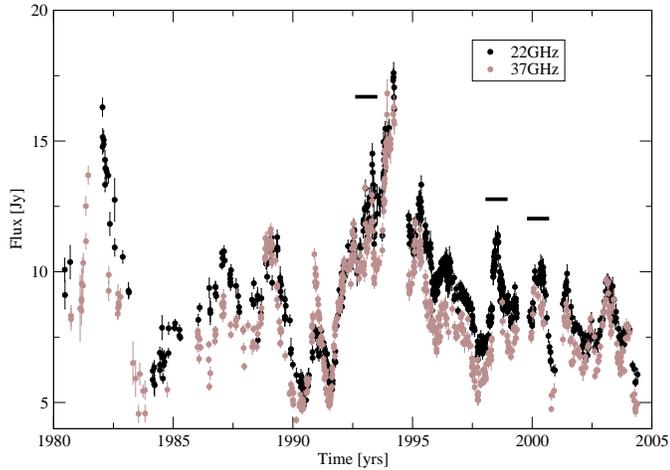}
   \caption{Single dish radio lightcurve of 3C\,454.3 at 37\,GHz and 22\,GHz taken at the Mets\"ahovi radio telescope. The timeranges of the ejection of the compontents are marked by the bars.  
           (\cite{ter})}
	   \label{fig:light}
    \end{figure}

 Figure \ref{fig:epochs} shows that in early 1994, the core appears elongated. This indicates the imminent ejection of a new
 component. Similar elongations and the ejection of new components can also be observed in 1997.28 and 1999.81. 
 In Figure \ref{fig:light} we show the flux density variability of 3C\,454.3 at 22\,GHz and 37\,GHz over the 
 last 25 years as measured by the Mets\"ahovi group (\cite{ter}, and Ter\"asranta priv. comm.).
 After the ejection of the new components, the flux increases at both frequencies but with a time delay of
 about eight weeks for the 37\,GHz flux and another four weeks also at 22\,GHz. 
 
 For two epochs, 1999.30 and 1999.81, we are able to compare the inner part of the jet of 3C\,454.3 from our 3\,mm VLBI maps with VLBI images obtained at
 7\,mm (Marscher, priv.comm). This is shown in Figure \ref{fig:7mm}.
 For the April data
 we derive a spectral index for the VLBI core of $\alpha$\,=\,0.64 (S$\sim\nu^{+\alpha}$), i.e. a strongly inverted spectrum. In 
 October the spectral index steepened to $\alpha$\,=\,--0.39 and a new component appeared west of the core
 (Fig.\ref{fig:epochs}). This new component is still nearly unresolved at 43\,GHz. Due to self-absorption of the jet increasing flux is first seen at high frequencies, in this case at 86\,GHz and later on at 43\,GHz.

  \subsection{An X-Ray view}
A comparably large amount of absorption towards the nuclear X-ray core of
3C\,454.3 of $N_{\rm H} = 5 \times 10^{21}$\,cm$^{-2}$ has been reported from a
{\it Beppo}Sax observation of this source (Tavecchio et al. \cite{Tav02}).
Alternatively, these authors suggest an intrinsic break in the continuum occurring below
$\sim$1\,keV to explain the observed lack of soft X-ray photons.
In both cases, important insights into the physics of AGN might be
provided by the combination of mm-VLBI radio observations on the smallest
accessible scales of the nuclear radio jet and the high-energy X-ray spectrum.
The most compact region of the nuclear radio jet represents an attractive candidate
for the dominating source of the compact X-ray emission in 3C\,454.3. In addition, the
angular resolution achieved in our 3mm-VLBI experiments of 50\,mas allows us to search
for hints of the putative compact absorber on linear scales as small
as 0.3\,pc. Such an approach is particularly interesting because of the puzzling
discrepancy between the expected small angle of the jet axis to the line of sight
in superluminal quasars like 3C\,454.3 and the occurrence of considerable absorption,
more naturally expected from type\,2 objects oriented closer to the line of sight.

We present here data from a short snapshot {\it CHANDRA} observation of 3C\,454.3
performed on Nov., 6th, 2002. This observation was done within the scope of a
survey of quasar jets (Marshall et al. \cite{Mar04}).Independent of that study, the X-ray properties of 3C\,454.3 were analysed as a part of an X-ray spectral survey of radio-loud core-dominated AGN (Kadler et al., these proceedings). The X-ray image (Fig.~4) reveals a resolved core-jet structure of 3C\,454.3 with
a bright knot of X-ray emission coinciding with strong radio emission emitted
about 5\,arcsec from the core (Murphy et al. 1993). In addition, a significant
unresolved source of X-ray emission is located at the same P.A. at a separation of
$\sim 1$\,arcmin from the core. Deeper large-scale radio imaging is necessary to
reveal the nature of this peculiar source.  The X-ray spectral analysis of the
integrated X-ray emission of 3C\,454.3 is in agreement with the results of
Marshall et al. (\cite{Mar04}): we find a strongly piled-up spectrum with an
intrinsic photon index of $\sim 1.3$ and a considerable amount of absorption of
$\sim 6 \times 10^{21}$\,cm$^{-2}$ confirming also the results of
Tavecchio et al. (\cite{Tav02}).

 \begin{figure}
   \centering
   \includegraphics[width=\columnwidth]{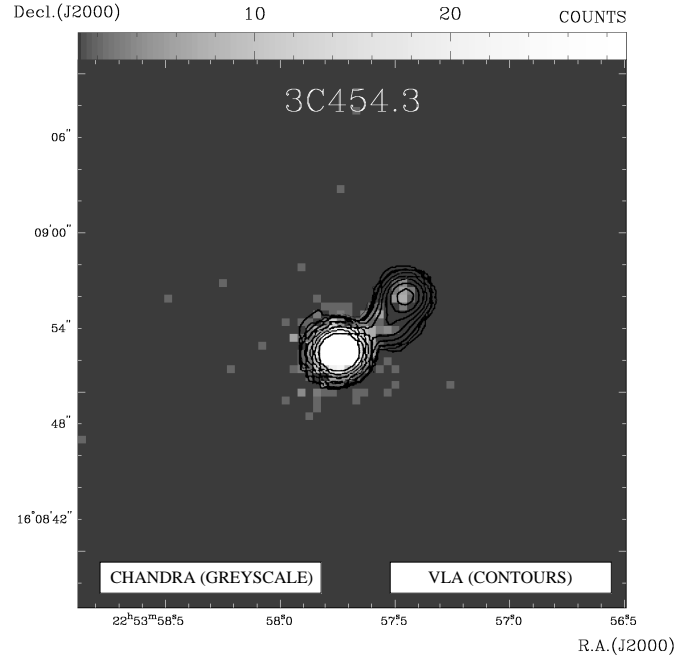}
   \caption{The kiloparsec-scale radio- and X-ray brightness distribution of 3C\,454.3. The
radio jet is displayed in contours (taken from Murphy et al. 1993) at a frequency
of 1.6\,GHz.}
	   \label{fig:xray}
    \end{figure}

\begin{acknowledgements}
3\,mm VLBI is a joint effort of the observatories listed in Table \ref{tab:antennas}. Without the help of many people at these
observatories this work would have not been possible. For communication of data prior to publication we also would like to thank S.
Jorstad, A. Marscher, and H. Ter\"asranta. M.\,Kadler were supported for this research through a stipend from the International Max Planck Research School (IMPRS) for Radio and Infrared Astronomy at the University of Bonn.

\end{acknowledgements}

\end{document}